\documentclass[a4paper, amsfonts, amssymb, amsmath, reprint, showkeys, nofootinbib, twoside, colorlinks=true, linkcolor=blue, citecolor=red, urlcolor=blue]{revtex4-2}

\usepackage[english]{babel}
\usepackage{esvect}
\usepackage{url}
\usepackage{orcidlink}
\usepackage[utf8]{inputenc}
\usepackage[colorinlistoftodos, color=green!40, prependcaption]{todonotes}

\usepackage{amsthm}
\usepackage{graphicx}
\usepackage[left=23mm,right=13mm,top=35mm,columnsep=15pt]{geometry}

\usepackage{adjustbox}
\usepackage{placeins}
\usepackage[T1]{fontenc}
\usepackage{csquotes}
\usepackage{changes}

\usepackage[normalem]{ulem}

\usepackage{booktabs}
\usepackage{amsmath}
\usepackage{hyperref}
\usepackage{bm} 
\usepackage{multirow}
\usepackage{xspace}

\def \teob{\textsc{TEOBResumS-Dali}\xspace}
\def \seob{\textsc{SEOBNRv5EHM}\xspace}

\def \bfteob{\textsc{6.454}\xspace}
\def \bfseob{$6.712$\xspace}
\def \bfteobHM{$7.814$\xspace}
\def \bfseobHM{$6.372$\xspace}
\def \bflogteob{$2.35$\xspace}

\def\apj{Astrophysical Journal}
\def\apjl{Astrophysical Journal Letters}
\def\apjs{Astrophysical Journal Supplement}

\def\mnras{Monthly Notices of the Royal Astronomical Society}

\def\prd{Physical Review D}
\def\prl{Physical Review Letters}
\def\pasa{Publications of the Astronomical Society of Australia}

\begin{document}
\title{Eccentricity signatures in LIGO-Virgo-KAGRA's Binary Neutron stars and Neutron-star Black Holes}

\author{{Keisi Kacanja$^{1}$}\orcidlink{0009-0004-9167-7769}}
    \email[kkacanja@syr.edu]{}

\author{{Kanchan Soni$^{1}$}\orcidlink{0000-0001-8051-7883}}
    \email[ksoni01@syr.edu]{}

\author{{Alexander H. Nitz$^{1}$}\orcidlink{0000-0002-1850-4587}}

\affiliation{$^1$Department of Physics, Syracuse University, Crouse Dr, Syracuse, NY 13210}

\date{\today}

\begin{abstract} 

Measurement of the eccentricity of low-mass binary systems through gravitational waves is crucial to distinguish between various formation channels. However, detecting eccentricity in these systems has been challenging due to the lack of accurate eccentric waveform models and the high computational cost associated with Bayesian inferences for systems with low-mass objects. In this work, we assess the eccentricities of seven previously observed low-mass gravitational wave events using publicly available data from the O1-O4a observing runs of the LIGO and Virgo observatories. We analyze the events using a new eccentric waveform model, \seob, and compare our results with the existing model \teob. We also present the first eccentricity constraints for GW190814. To improve the accuracy of our parameter estimation, we include higher-order modes in both waveform models. We optimize inference by employing efficient marginalization techniques to alleviate the computational costs associated with low-mass systems, and parallelization techniques for sampling large parameter spaces. We find that one of the sources, GW200105, exhibits a non-negligible eccentricity, with a measured eccentricity of $e=0.125^{+0.029}_{-0.082}$ at 20 Hz ($90\%$ credible level) for \seob and $e=0.135^{+0.019}_{-0.088}$ for \teob for a uniform eccentricity prior ranging from 0 to 0.2 at a reference frequency of 20 Hz. We find moderate support for the eccentric waveform hypothesis with a Bayes factor of $\sim 6-7$ times more preference for the eccentric model over the non-eccentric one. When using a uniform log prior on eccentricity with a minimum bound of $10^{-4}$, the support for the eccentric model decreases, with the Bayes factor reduced to \bflogteob. For the remaining five sources, the results are consistent with low eccentricity, with 90\% upper limits ranging from $e \leq 0.011$ to $e \leq 0.066$. We do not find any support for non-negligible eccentricity in GW190814. Finally, we discuss the challenges of performing Bayesian inference in eccentric, multi-modal parameter spaces, including issues related to sampling efficiency and waveform systematics. 
\end{abstract}


\maketitle

\section{Introduction} \label{sec:intro}
 
Advanced LIGO and Advanced Virgo~\citep{advancedligo,advancedvirgo} have detected hundreds of gravitational wave (GW) signals from the mergers of compact objects during their third observing run (O3), covering a wide range of source masses~\citep{gwtc3,ogc4_Nitz_2023,venumadhav_2022_o3a,mehta2024newbinaryblackhole,wadekar2023newblackholemergers}. Although the majority of these events are binary black hole (BBH) mergers, a handful of low-mass events fall within the binary neutron star (BNS) and neutron star–black hole (NSBH) mass ranges. These include GW170817~\citep{gw170817}, GW190425~\citep{GW190425} identified as BNS mergers, and GW200105,  GW200115~\citep{nsbh_ligo}, and the two recently observed objects GW230529 and GW230518 ~\citep{gw230529,4gwtc,LIGOScientific:2025pvj} in the first part of the fourth observing run (O4a) of LIGO, identified as NSBH mergers based primarily on their component masses. These detections have enabled observations and constraints on several known features of compact binary systems, such as the spin-induced precession of the binary’s orbit in GW190412~\citep{GW190412, Hannam:2021pit, LIGOScientific:2025brd} and the presence of subdominant harmonics in GW190412 and GW190814~\citep{GW190412,gw190814}. 

Until the end of O3, there was no compelling observational evidence for orbital eccentricity in compact binary mergers in low-mass systems such as NSBHs, BNSs, or even low-mass BBHs. However, recent analyses of GW200105 have indicated possible hints of eccentricity, marking it as the first NSBH event with potential eccentricity~\citep{Fei_2024, 2025arXiv250315393M,planas2025eccentricinspiralmergerringdownanalysisneutron}. This observation challenges the assumption of an isolated binary origin and suggests a formation channel beyond standard binary evolution models~\citep{Zevin_2020,Zevin_2021,2gw190814_formation}.

Eccentricity leaves a characteristic imprint on the gravitational-wave signal by modulating the phase evolution during the early inspiral. As the binary orbits, the orbital eccentricity dissipates rapidly and generally circularizes as the system approaches the merger phase unless acted upon by external forces~\citep{PhysRev.136.B1224}. However, if residual eccentricity persists in the frequency band of ground-based detectors like LIGO and Virgo (around $10-20$ Hz), it can be detectable. Since the binary’s orbital eccentricity carries traces of its evolutionary history, measuring it in a GW signal can serve as a smoking gun indicator for its distinct formation channel~\citep{Zevin_2021}.

A compact binary system can exhibit measurable orbital eccentricity if it is either born with high eccentricity or acquires it through external interactions over time. The first scenario is more common for isolated stellar-mass binary stars that evolve through pathways such as the common envelope phase~\citep{1976IAUS...73...75P,Livio_1996,Belczynski_2002,Dominik_2012}, chemically homogeneous evolution~\citep{Mandel_2016,10.1093/mnras/stw1219}, or fallback-driven mechanisms in failed supernovae~\citep{2018PhRvL.120z1101T}. In these cases, eccentricity is unlikely to be detectable, as it is expected to dampen to $e \lesssim 10^{-3}$ when the GW signal emitted enters the LIGO-Virgo frequency band ($> 10\, \rm Hz$)~\citep{Kowalska_2011,Mapelli:2020vfa}. The second scenario typically occurs in dense environments, such as nuclear star clusters, young star clusters, and globular clusters, in which binaries form and harden through dynamical interactions ~\citep{RevModPhys.50.437,1993Natur.364..423S}. In these regions, dynamical interactions, such as frequent gravitational encounters and dynamic friction~\citep{Gondan:2017wzd,Sedda:2023qlx}, can induce measurable eccentricity in the orbits of compact binaries.

Although dynamical formation channels predict large eccentricities for high-mass binaries,theoretical studies generally find that the overall fraction of binaries with measurable eccentricity is small, typically a few percent, and even lower for low-mass systems~\cite{Samsing:2017xmd,Zevin:2018kzq,Tagawa:2020jnc}. This makes the observation of an eccentric low-mass binary even more unlikely ~\citep{2019MNRAS.490.5210R,Wu_2020,2020arXiv200905461G,2021MNRAS.506.1665G,Romero_Shaw_2021,2023PhRvD.107f4024B,2024ApJ...972...65I}. Nevertheless, there has been growing interest in exploring eccentricity in low-mass systems as well, due to the recent support for eccentricity in GW200105 \citep{Fei_2024,2025arXiv250315393M,planas2025eccentricinspiralmergerringdownanalysisneutron}. However, inferring properties from these systems is challenging due to several computational challenges. First, including eccentricity into the search space significantly increases dimensionality and parameter ranges, making matched filtering searches much more computationally expensive~\citep{PhysRevD.110.044013,Phukon:2024amh,dhurkunde_nitz_2025}. Second, eccentricity is most pronounced during the early inspiral stages, where the LIGO and Virgo detector's poor low-frequency sensitivity ($<10  \, \rm Hz$) limits their detectability. Third, searching for these features at lower frequencies extends the signal duration, further increasing the computational cost of Bayesian inference. Finally, the challenge is compounded by the lack of accurate waveform models, making such searches or Bayesian inferences even more difficult.

In this study, we present a systematic investigation of orbital eccentricity in low-mass gravitational-wave events observed by LIGO and Virgo between O1 and O4a. Our analysis focuses on inferring eccentricity using the \seob model, an effective-one-body (EOB) formalism~\citep{PhysRevD.59.084006,PhysRevD.62.064015} built over \textsc{SEOBNRv5} waveform family~\citep{Khalil:2023kep,Ramos-Buades:2023ehm,Pompili:2023tna,vandeMeent:2023ols,Mihaylov:2023bkc} that incorporates higher-order modes and eccentricity effects \citep{seob,Gamboa:2024hli}. For comparison, we also use the \teob model~\citep{2024PhRvD.110h4001N} as an alternative EOB-based waveform model to assess the consistency and accuracy of eccentricity constraints across models. Our analysis is performed over the dominant (2,2) mode of the waveform, as well as the higher-order modes (HOM) up to $l =4$, which include subdominant multi-polar contributions beyond the quadrupole that become important in systems such as high mass ratios or eccentric orbits.

We find that six out of the seven events are consistent with circular orbits, placing stringent upper bounds on eccentricity. However, for GW200105, we observe moderate support for non-zero eccentricity across both waveform models. In particular, GW200105 exhibits a well-constrained eccentricity of $0.125^{+0.029}_{-0.082}$ for \seob and $0.135^{+0.019}_{-0.088}$ for \teob at the 90\% credible level, including HOMs and a reference frequency of 20 Hz, consistent with previous studies~\citep{Fei_2024, 2025arXiv250315393M, planas2025eccentricinspiralmergerringdownanalysisneutron}.

To better understand the underlying structure of the posterior distributions and the sampling challenges, we also conduct a thorough investigation of the eccentricity–anomaly parameter space. Our analysis reveals multi-modal features and isolated regions in the posterior distributions, particularly for systems with non-negligible eccentricity such as GW200105. We perform multiple checks, including varying the frequency cutoff across multiple inferences, injecting signals at higher frequency values, calculating the likelihoods within the peaks and dips of the modes, and using alternate samplers. Furthermore, by directly computing waveform overlaps while varying only the eccentricity, we demonstrate that these modes arise from the waveform model itself and not from the sampling procedure. This behavior is consistent in both \seob and \teob and suggests that the observed structure is physical and not due to waveform systematics.

\section{Method}

We assess the eccentricity of low-mass gravitational wave events by performing Bayesian inference~\citep{1763RSPT...53..370B,2006CQGra..23.4895R} using the \textsc{PyCBC Inference} toolkit~\citep{2019PASP..131b4503B} and obtain the posterior probability distribution $p(\vec{\theta}|d,h)$ defined as  

\begin{equation}
    p(\vec{\theta}|d,h) = \frac{\mathcal{L}(d|\vec{\theta},h) \pi(\vec{\theta}|h)}{\mathcal{Z}}\,. 
    \end{equation}

The posterior distribution represents the probability of the parameters $\vec{\theta}$ given the observed data $d$ and the model $h$ \citep{Veitch_2010,2019PASA...36...10T}. The likelihood $\mathcal{L}(d|\vec{\theta},h)$ quantifies how well the data match a particular set of parameters and the waveform model. $\pi(\vec{\theta}|h)$ is the prior distribution for a set of parameters. The prior represents our initial assumptions or beliefs about the ranges of the parameters, based on existing knowledge or theoretical considerations. $\mathcal{Z}$ is the evidence and serves as a normalization factor in this case. $\vec{\theta}$ represents the intrinsic and extrinsic parameters of the binary system. Intrinsic parameters include the masses of the compact objects, their spins, the tidal deformability of the compact object if it is a neutron star, the orbital eccentricity, and the radial phase parameter, which characterizes the binary's relative position within a Keplerian orbit. The radial phase can be expressed as either the true anomaly or the relativistic anomaly. The extrinsic parameters consist of the distance from the source, its location, the angle of inclination to the detector, the angle of polarization with respect to the detector, and the time and phase of the binary coalescence.

\subsection{Waveform Model}

To measure eccentricity in low-mass gravitational-wave events, we employ two time-domain waveform models within the EOB framework: \seob~\citep{seob_theory,seob} and \teob~\citep{teob}. Both models generate HOM eccentric inspiral–merger–ringdown waveforms and are calibrated with numerical relativity simulations. The \seob model is built for binaries with eccentric orbits and spins aligned (or antialigned) with the orbital angular momentum. In addition to the dominant (2,2) mode, it includes subdominant modes such as (2,1), (3,3), (3,2), (4,4), and (4,3), allowing for more accurate modeling of systems with mass asymmetry or orbital inclination. It is important to note that this model utilizes the definition of relativistic anomaly to describe the position of the binary in an eccentric orbit.

\teob is another time-domain waveform model with the ability to model a wide range of compact binary systems, including BNS, BBH, and NSBH binaries~\citep{teob}. It incorporates a comprehensive set of physical effects critical for accurate waveform modeling, including orbital eccentricity, hyperbolic encounters, non-planar orbital dynamics, spin-orbit and spin-spin couplings, and the tidal deformability of neutron stars. Additionally, the model includes HOMs up to multipole order $l = 8$. Instead of the relativistic anomaly, \teob uses the true anomaly definition to describe the radial phase of the eccentric orbit. 

Since neither model is originally formulated in the frequency domain, we apply a Fast Fourier Transform with appropriate tapering to convert the waveform for use in our frequency-domain inference pipeline. Before performing the inference, we validate that the waveform was generated successfully across our parameter space and that all tested waveforms were produced without issues.

\subsection{Data and Priors}

We analyze seven events---GW170817, GW190425, GW190814, GW200105, GW200115, GW230518, and GW230529 using publicly available strain data from the LIGO Hanford, LIGO Livingston, and Virgo detectors~\citep{GWOSC1,GWOSC2,gwosclink}. For each event, we include data only from the detectors that were online at the time of the observation. GW190425 and GW200105 were recorded with the Livingston-Virgo network, GW235018~\cite{4gwtc,LIGOScientific:2025pvj} with the Hanford-Livingston network, and GW230529 was observed only by LIGO Livingston. Although both Livingston and Virgo data were available for GW200105, we excluded Virgo from our analysis. Virgo's lower sensitivity and unfavorable sky location significantly reduce its contribution to the signal, particularly in the lower frequency band relevant for an eccentric study. For computational efficiency and to minimize excessive artifacts, we only use the Livingston data in our final analysis for GW200105. 

We choose prior bounds that align with the astrophysical properties measured in the Fourth Open Gravitational Wave Catalog~\citep{ogc4_Nitz_2023} and the Fourth Observation Run Discovery Paper, except for precessing spins~\citep{2024ApJ...970L..34A}. Specifically, we choose uniform priors over the detector frame chirp mass $\mathcal{M}$, the mass ratio $q$, and eccentricity $e$. We choose the bounds on the eccentricity to range from 0 to 0.2 for GW200105 and from 0 to 0.1 for the other sources. Additionally, we perform a separate inference using a uniform-in-log prior on eccentricity for GW200105 to assess the dependence of our results on the prior choice. For this run, we choose a minimum bound of $10^{-4}$ and an upper bound of 0.2. The reference frequency is set to be equal to the gravitational wave frequency of 20 Hz. Incorporating eccentricity, HOMs, and precession into gravitational waveform models remains an area of ongoing development, and current models have varying levels of support for these effects. Due to existing model limitations, we adopt a broad agnostic prior with uniform aligned spins, $\chi_{1z}$ and $\chi_{2z}$. We choose uniform angle priors for the polarization $\Psi$, relativistic anomaly $\ell_\text{rel}$ for \seob and true anomaly $\ell_t$ for \teob. We choose a uniform luminosity volume prior for $d_{L}$. We assumed an isotropic sky distribution for the inclination $\iota$, right ascension $\alpha$, and declination $\delta$. In our study, we do not consider the tidal deformability of neutron stars. The reason is that we do not expect the tidal effects to be measurable with the current detector's sensitivity for the given events. As a result, its inclusion is unlikely to affect the measurement of the eccentricity parameter.

\subsection{Parameter Estimation Optimizations}

Bayesian inference for low-mass binaries is computationally demanding, primarily due to the long-duration signals and the slow waveform generation associated with EOB models. For example, generating a waveform from our highest likelihood model for GW190425 took 12.3 seconds to generate in the frequency domain. During parameter estimation, $\sim\mathcal{O}(10^6)$ likelihoods are calculated, making parameter estimation for these systems especially costly. Although reduced order models are often used in analyses to speed up EOB waveform evaluation, no such surrogate currently exists for the parameter space and models we utilized~\citep{rom2,rom3,rom1}. As a result, we directly use the complete EOB models, which require the numerical solution of coupled ordinary differential equations to compute the waveform dynamics~\citep{SEOBNRv5,teob}. This waveform generation cost is further amplified when modeling complex systems, such as eccentric orbits with HOMs. In addition, the high dimensionality and complex structure of the parameter space substantially increase the computational cost of efficient sampling and can lead to slow convergence or inefficient exploration of the posterior. 

Given that the computational cost of Bayesian inference is expected to be large for these signals, we employ a few optimizations. One optimization includes reducing the dimensionality of our parameter space to make sampling more efficient. To do this, we employ marginalization techniques~\citep{2019PASP..131b4503B,importance_sampling} in which we integrate the time of coalescence, sky location, distance, and polarization angle from the likelihood calculation. This approach improves sampling efficiency and convergence, thereby accelerating the inference process. Marginalization over distance is performed using a precomputed lookup table, while sky location, time of arrival, and polarization are sampled from a proposal distribution approximating the posterior~\citep{importance_sampling,importance_sampling2}. The samples are then reweighted by the ratio of the posterior to the proposal, following the method outlined in Ref.~\cite{importance_sampling}. Marginalization over the coalescence phase is also feasible. However, this method can only be applied to the dominant (2,2) mode waveforms, as marginalization techniques over the coalescence phase are not currently implemented for waveforms with HOMs and remain an ongoing area of development. As a result, the phase must be explicitly sampled, increasing the dimensionality of the parameter space and further adding to the computational costs.

For events with non-negligible eccentricity, we also utilize a parallelization strategy by splitting the prior range of eccentricity evenly into multiple bins, running independent inference in each, and recombining in postprocessing. This approach helps mitigate the complexity of the likelihood surface: eccentric binaries introduce strong, nonlinear structure in the posterior, which can be highly multi-modal and slow to converge when explored globally. By dividing the prior into narrower intervals, each sampler explores a smaller, more manageable region of parameter space, improving convergence and reducing the overall computational cost.

To efficiently explore our high-dimensional parameter space, we utilize the nested sampler \textsc{Dynesty}~\citep{2020MNRAS.493.3132S}. Nested sampling is particularly well-suited for handling complex, multi-modal distributions and efficiently sampling high-dimensional likelihood surfaces. This is especially beneficial in our analysis, where the incorporation of both eccentricity and HOMs significantly increases the complexity of the likelihood function.

\section{Observations}  

We present the first eccentricity constraints for seven low-mass gravitational-wave events using the newly developed \seob waveform model. To ensure robustness, we cross-validate our results using the \teob model, which offers a complementary treatment of eccentricity.

\begin{figure*}
    \centering
    \includegraphics[width=\linewidth]{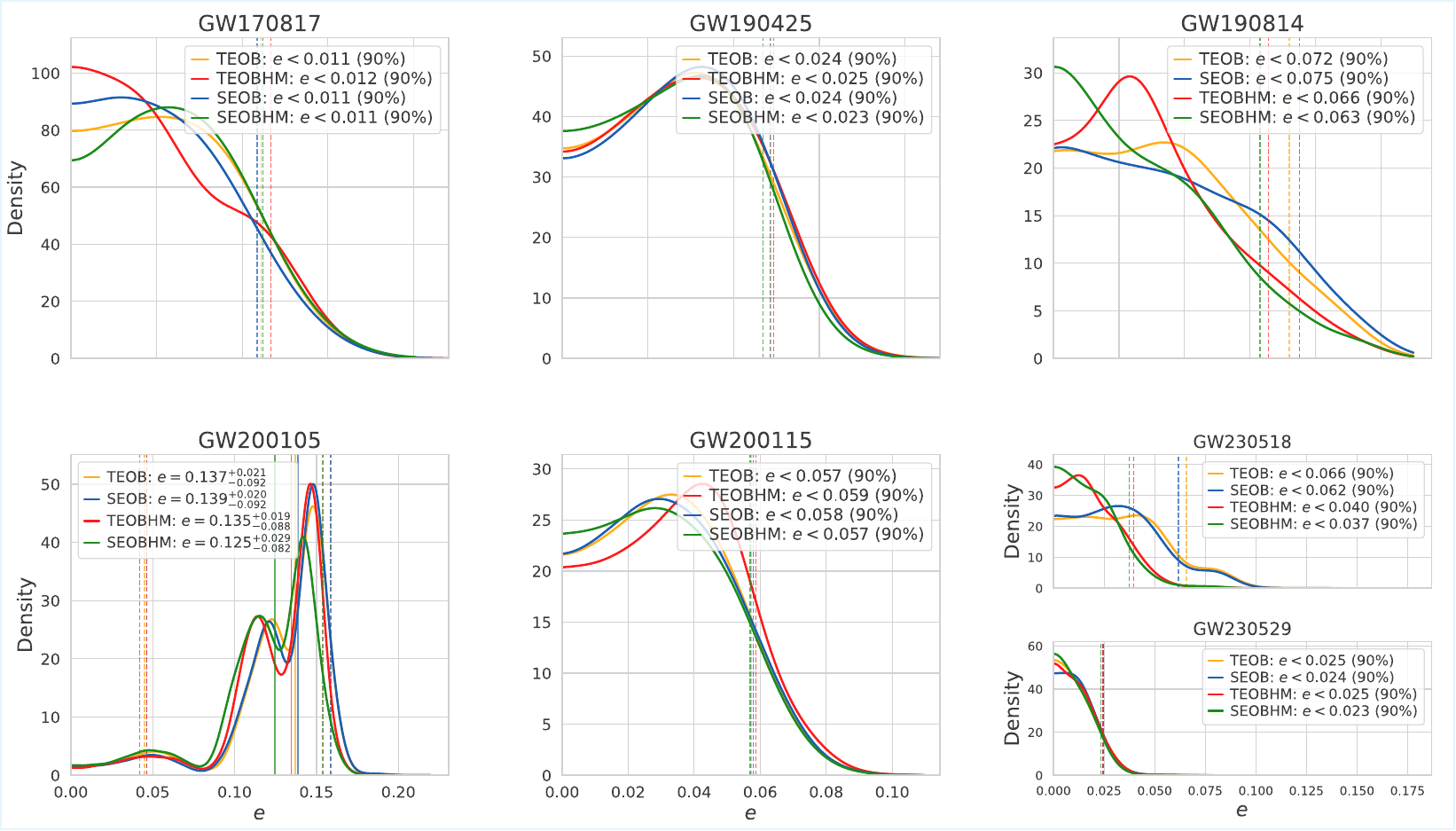}
    \caption{Normalized one-dimensional posterior distributions for the orbital eccentricity $e$ at 20 Hz for seven low-mass gravitational wave events, using \seob and \teob depicted in blue and orange for the dominant (2,2) waveform mode (SEOB,TEOB) and with HOMs included depicted in green and red (SEOBHM, TEOBHM). The posteriors have been smoothed using a Gaussian kernel density estimate. The dotted vertical line indicates the bounds of the 90\% confidence interval in each panel. For GW200105, the dashed lines represent the 90\% credible interval.}
    \label{fig:ecc}
\end{figure*}

\subsection{GW170817, GW190425, GW190814, GW200115, GW230518 and GW230529}

Our results for GW170817, GW190425, GW200115, and GW230529 are consistent with previous parameter estimation analyses~\citep{ogc4_Nitz_2023, Lenon_2020, 2020MNRAS.496L..64R, Fei_2024, 2024ApJ...970L..34A}. The most probable parameters for each event are summarized in Table~\ref{tab:results}, and the corresponding eccentric posteriors are shown in Fig.~\ref{fig:ecc}.

For GW170817, we constrain the orbital eccentricity to $e < 0.011$ at a reference frequency of 20 Hz. This estimate is consistent with the upper limits reported in~\citet{Lenon_2020} which report a bound of $e<0.024$ at a gravitational wave frequency of 10 Hz, \citet{PhysRevD.110.042003} with a bound of $e<0.01$  at 20 Hz, and~\citet{Nitz_ecc_2020} with a bound of $e<0.02$ at a reference frequency of 30 Hz. For GW190425, we find $e < 0.013$ at 20 Hz. To unify the eccentric definitions, we calculate the equivalent eccentricity for a reference frequency of 10 Hz by calculating a minimizing mismatch function~\citep{ecc_defs}. With a mismatch of 0.0036 in comparison to our 20 Hz reference, we obtain the eccentricity to be 0.0464, which is comparable to previous constraints: $e \leq 0.048$ at 10 Hz reported in~\citet{Lenon_2020}, and $e \leq 0.007$ at 10 Hz obtained using the \textsc{SEOBNRE} model in~\citet{Romero_Shaw_2020_GW190425,SEOBNRE}.

Our recovery of mass estimates for GW190814 is consistent with previous analyses~\citep{ogc4_Nitz_2023}. We constrain the eccentricity to be $e< 0.063$ for the HOM run using \seob. However, we observe that the \teob model shows a mild peak near $e\sim0.03$ in the HOM run, which is not present in the corresponding \seob result. Although differences between the dominant (2,2) mode and HOM runs can be attributed to the increased sensitivity of higher-order modes, particularly relevant for GW190814, a system known to exhibit strong HOMs, the discrepancy between \seob and \teob in the HOM runs may reflect waveform systematics. In particular, GW190814’s large mass ratio presents a significant modeling challenge, and waveform inaccuracies at such extreme mass ratios could lead to differences in the inferred eccentricity distributions, as seen in Fig.~\ref{fig:ecc}.

For GW200115, we place an upper limit of $e < 0.024$ for a reference frequency of 20 Hz, broadly consistent with the constraint of $e<0.02$ also for a reference frequency of 20 Hz reported in~\citet{planas2025eccentricinspiralmergerringdownanalysisneutron} and the first constraint of $e < 0.06$ at 10 Hz reported in~\citet{Fei_2024}. However, our eccentric posterior distribution is not bimodal, unlike~\citet{Fei_2024}. For GW230529, we find that our constraint of $e < 0.023$ for a reference frequency of 20 Hz using both waveform models is consistent with~\citet{planas2025eccentricinspiralmergerringdownanalysisneutron} where they place a bound of $e<0.02$ for the same frequency reference. For GW230518, we find no evidence of eccentricity and place an upper bound of $e<0.04$ for the HOM models.

Overall, we find no significant evidence for eccentricity in these previously detected BNS and NSBH systems, with the exception of GW200105.

\begin{table*}[!]
    \centering
\caption{Measured eccentricities and source parameters for low-mass gravitational-wave events analyzed using the \teob and \seob models, both with and without the inclusion of HOM. The parameters include eccentricity $e$, detector-frame chirp mass $\mathcal{M}$, and mass ratio $q$ $(m_1/m_2\,, m_1 > m_2)$. For eccentricity, we report the 90\% upper limit if the 90th percentile is below 0.1; otherwise, we report the median with symmetric 90\% credible intervals. For GW200105, we always report the median with symmetric 90\% credible intervals.}
    \label{tab:results}
    \setlength{\tabcolsep}{6pt}
    \renewcommand{\arraystretch}{1.2}
    \begin{tabular}{lccccccc}
    	\toprule
    	\multirow{2}{*}{\textbf{Event}} & \multicolumn{3}{c}{\textsc{TEOBResumS-Dali}} & & \multicolumn{3}{c}{\textsc{SEOBNRv5EHM}} \\
    	\cmidrule{2-4} \cmidrule{6-8}
    	& \textbf{e} & \bm{$\mathcal{M}$} & \bm{$q$} & & \textbf{e} & \bm{$\mathcal{M}$} & \bm{$q$} \\
    	\midrule
    	GW170817 & $<0.011$ & $1.19750^{+0.00014}_{-0.00019}$ & $1.264^{+0.215}_{-0.236}$ & & $<0.011$ & $1.19752^{+0.00013}_{-0.00019}$ & $1.231^{+0.233}_{-0.211}$ \\
    	\textbf{GW170817 (HOM)} & $<0.012$ & $1.19752^{+0.00015}_{-0.00021}$ & $1.323^{+0.157}_{-0.278}$ & & $<0.011$ & $1.19751^{+0.00013}_{-0.00018}$ & $1.311^{+0.163}_{-0.266}$ \\
    	\midrule
    	GW190425 & $<0.024$ & $1.48632^{+0.00064}_{-0.00103}$ & $1.327^{+0.519}_{-0.299}$ & & $<0.024$ & $1.48627^{+0.00056}_{-0.00089}$ & $1.241^{+0.387}_{-0.218}$ \\
    	\textbf{GW190425 (HOM)} & $<0.025$ & $1.48633^{+0.00064}_{-0.00105}$ & $1.358^{+0.547}_{-0.331}$ & & $<0.023$ & $1.48635^{+0.00066}_{-0.00090}$ & $1.308^{+0.528}_{-0.279}$ \\
    	\midrule
    	GW190814 & $<0.072$ & $6.44146^{+0.04976}_{-0.08109}$ & $11.710^{+3.032}_{-5.257}$ & & $<0.075$ & $6.43642^{+0.05107}_{-0.07938}$ & $11.489^{+3.195}_{-5.119}$ \\
    	\textbf{GW190814 (HOM)} & $<0.066$ & $6.43014^{+0.05224}_{-0.06175}$ & $10.279^{+3.264}_{-3.255}$ & & $<0.063$ & $6.42758^{+0.04637}_{-0.05842}$ & $9.970^{+3.452}_{-2.751}$ \\
    	\midrule
    	GW200105 & $0.137^{+0.021}_{-0.092}$ & $3.56297^{+0.04092}_{-0.02674}$ & $3.767^{+0.931}_{-2.339}$ & & $0.139^{+0.020}_{-0.092}$ & $3.56483^{+0.04146}_{-0.02750}$ & $3.813^{+0.973}_{-1.203}$ \\
    	\textbf{GW200105 (HOM)} & $0.135^{+0.019}_{-0.088}$ & $3.56845^{+0.03722}_{-0.02784}$ & $3.681^{+1.128}_{-2.009}$ & & $0.125^{+0.029}_{-0.082}$ & $3.57608^{+0.03388}_{-0.03243}$ & $3.781^{+1.131}_{-2.213}$ \\
    	\midrule
    	GW200115 & $<0.057$ & $2.57372^{+0.00901}_{-0.01144}$ & $2.744^{+3.175}_{-0.690}$ & & $<0.058$ & $2.57235^{+0.00957}_{-0.01076}$ & $2.196^{+3.347}_{-0.790}$ \\
    	\textbf{GW200115 (HOM)} & $<0.059$ & $2.57124^{+0.00836}_{-0.01031}$ & $2.113^{+2.702}_{-0.680}$ & & $<0.057$ & $2.57239^{+0.00852}_{-0.01039}$ & $2.147^{+3.252}_{-0.735}$ \\
    	\midrule
    	GW230518 & $<0.066$ & $2.93215^{+0.01504}_{-0.02181}$ & $3.813^{+3.274}_{-2.220}$ & & $<0.062$ & $2.93356^{+0.01378}_{-0.02178}$ & $4.134^{+2.863}_{-2.460}$ \\
    	\textbf{GW230518 (HOM)} & $<0.040$ & $2.94097^{+0.00917}_{-0.01372}$ & $5.368^{+2.343}_{-2.335}$ & & $<0.037$ & $2.94101^{+0.00877}_{-0.01327}$ & $5.306^{+2.363}_{-2.217}$ \\
    	\midrule
    	GW230529 & $<0.025$ & $2.02437^{+0.00282}_{-0.00211}$ & $1.604^{+1.919}_{-0.564}$ & & $<0.024$ & $2.02449^{+0.00319}_{-0.00215}$ & $1.696^{+2.233}_{-0.644}$ \\
    	\textbf{GW230529 (HOM)} & $<0.025$ & $2.02429^{+0.00258}_{-0.00207}$ & $1.502^{+1.958}_{-0.457}$ & & $<0.023$ & $2.02432^{+0.00266}_{-0.00181}$ & $1.479^{+1.777}_{-0.439}$ \\

    	\bottomrule
    \end{tabular}
\end{table*}

\subsection{GW200105}

The first evidence for eccentric features in GW200105 was reported by~\citet{Fei_2024}, who measured an eccentricity of approximately $0.07^{+0.01}_{-0.01}$ at a reference frequency of 20 Hz using a modified version of the \textsc{IMRPhenomXPHM} approximant~\citep{SEOBNRE}. Since then, similar studies have estimated the eccentricity parameter for GW200105, further supporting the presence of eccentricity in the signal. For example, \citet{2025arXiv250315393M} reported a median eccentricity of $\sim0.145^{+0.007}_{-0.097}$ ($90\%$ credible interval) using several waveform models including \textsc{TaylorF2Ecc}, \textsc{pyEFPE}, \textsc{IMRPhenomXP}, and \textsc{IMRPhenomXPHM}~\citep{taylorf2ecc,pyEFPE,IMRPhenomXP_1,IMRPhenomXP_2}, evaluated at a reference frequency of 20 Hz. More recently,~\citet{planas2025eccentricinspiralmergerringdownanalysisneutron} measured the eccentricity as $\le 0.12^{+0.02}_{-0.03}$ for a reference frequency of 20 Hz using the waveform model \textsc{IMRPhenomTEHM}~\citep{IMRPhenomTEHM}. In this work, we measure the eccentricity of GW200105 to be $e_\text{SEOB} = 0.139^{+0.020}_{-0.092}$ using \seob and $e_\text{TEOB} =  0.137^{+0.021}_{-0.092}$ using \teob, both evaluated at a reference frequency of 20 Hz. When including HOMs, we obtain $e_\text{SEOB} = 0.126^{+0.027}_{-0.082}$ and $e_\text{TEOB} = 0.135^{+0.019}_{-0.088}$. While the inclusion of HOMs slightly reduces the eccentricity estimate in \seob, the distribution remains broadly consistent between both waveform models. 

\begin{figure*}
    \centering
    \includegraphics[width=\textwidth]{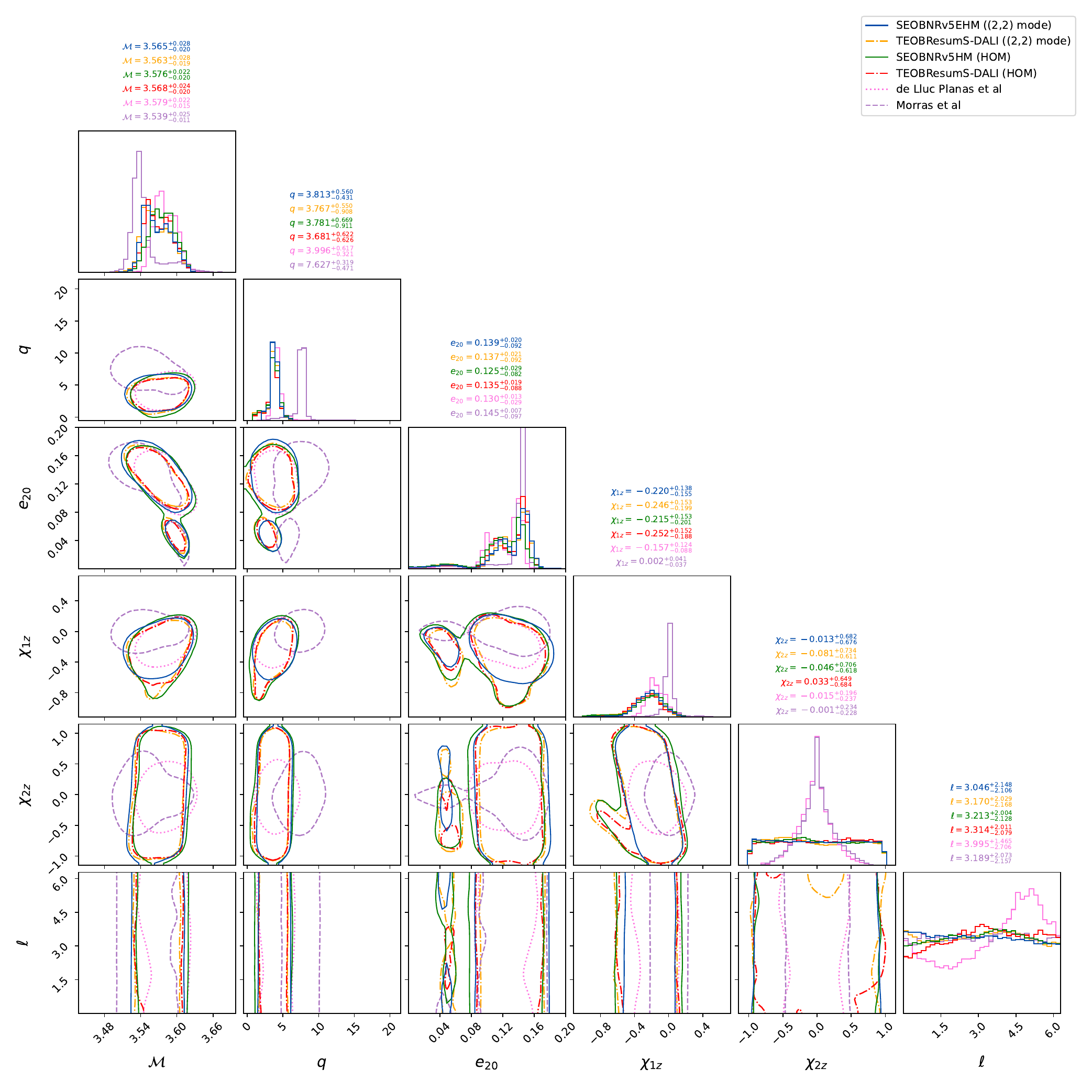}
    \caption{Corner plot comparing marginalized posterior distributions for GW200105. The parameters shown are chirp mass $\mathcal{M}$, mass ratio $q = m_1/m_2$ where $m_1,m_2$ are the masses of the primary and secondary object, eccentricity $e$ defined at 20 Hz, aligned spin components $\chi_{1z}$ and $\chi_{2z}$, and mean or relativistic anomaly $\ell$ which is different depending on the model. The distributions are plotted for dominant modes of \seob (dominant mode green solid line), \seob (HOM blue solid line), \teob (dominant mode yellow dash-dot line), and \teob (HOM red dash-dot line). The two-dimensional panels display contours representing the 90\% credible regions. The one-dimensional histograms along the diagonal show the marginalized posterior distributions for chirp mass, mass ratio, eccentricity, spins, and anomaly, with the quoted numbers representing the 90\% credible interval. For additional comparison, published results from~\citet{2025arXiv250315393M} using \textsc{pyEFPE} (purple dashed line) and~\citet{planas2025eccentricinspiralmergerringdownanalysisneutron} using \textsc{IMRPhenomTEHM} (pink dotted line) are also included.~\citet{planas2025eccentricinspiralmergerringdownanalysisneutron} reports the only results with no second island on the eccentricity and chirp mass or mass ratio posteriors, or a mild peak near $e\sim0.05$. \citet{2025arXiv250315393M} reports the only result with no secondary peak $e \sim 0.12$ in the posterior distribution.}
    \label{fig:comparison}
\end{figure*}

Our findings are presented in Fig.~\ref{fig:comparison}, where all previous results, except \citet{Fei_2024}, have been overlaid. All these measurements support the presence of non-zero eccentricity in GW200105 under the assumption of a uniform prior on eccentricity, suggesting the system may not have fully circularized before merger. However, some support still exists near zero, leaving open the possibility of a quasicircular origin.

\begin{figure}[h]
    \centering
    \includegraphics[width=0.48\textwidth]{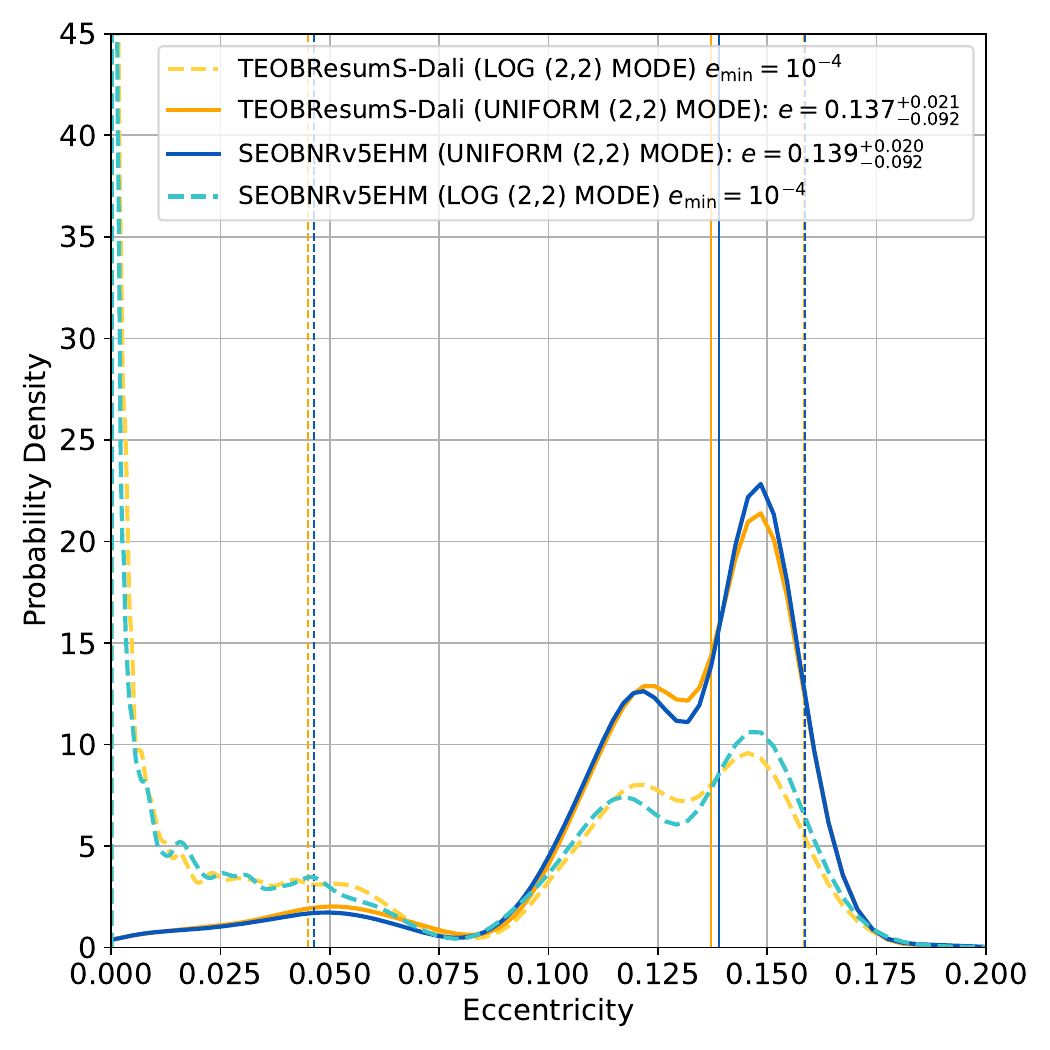}\hspace{0.06\textwidth}
    \caption{Comparison of eccentricity posteriors for GW200105 using different priors. None of the waveforms include HOMs in the plots. The orange curve shows the posterior assuming a uniform prior on eccentricity. The yellow dashed curve corresponds to a posterior assuming a log-uniform prior, with a minimum bound of $e_{\mathrm{min}} = 10^{-4}$. The dark blue curve represents the posterior on eccentricity with a uniform prior using \seob, while the dashed dark blue line shows the corresponding posterior under a log-uniform prior with the same lower bound. Vertical dashed lines indicate the median and 90\% credible intervals for each corresponding uniform distribution. The 90\% credible intervals are depicted in the legend.}
    \label{fig:log}
\end{figure}

To further assess the impact of prior choice, we repeated our inference using a prior uniform in $\log_{10}(e)$, with a lower bound of $10^{-4}$. This prior strongly favors lower eccentricities and shifts posterior support accordingly. While the peak near moderate eccentricity persists, the distribution becomes bimodal, with increased support at low eccentricity. This outcome highlights the critical influence of the prior in eccentricity inference, especially when the likelihood surface is broad or multi-modal. The corresponding posterior distributions under both priors are shown in Fig.~\ref{fig:log}, emphasizing the need for careful prior specification when interpreting eccentric gravitational-wave signals.

While our eccentricity estimates generally align with previous studies, some differences remain. We observe three distinct peaks in our eccentricity posteriors, whereas other works recover two. In particular,~\citet{planas2025eccentricinspiralmergerringdownanalysisneutron} does not observe the slight peak near $e\sim0.05$ eccentricity seen in our results and those of~\citet{2025arXiv250315393M}. Additionally,~\citet{2025arXiv250315393M} does not recover the dual peak structure that we, along with~\citet{planas2025eccentricinspiralmergerringdownanalysisneutron}, observe near $e \sim 0.12$. Both \seob and \teob agree very well with each other and recover the triple peak distribution. The detector-frame chirp masses are consistent across studies, indicating agreement on the overall signal scale. However, our inferred mass ratio differs significantly from~\citet{2025arXiv250315393M}, who measure $q = 7.63^{+0.32}_{-0.47}$, compared to our \seob HOM estimate of $q =3.574^{+0.035}_{-0.030}$. These discrepancies may stem from differences in waveform systematics across multiple waveforms.

Since both our models are non-precessing and assume aligned spins, it does not capture the full spin dynamics that may be present in the signal, unlike the models employed by~\citet{2025arXiv250315393M}, though this model is inspiral only and does not capture the full waveform dynamics. These differences highlight the impact of modeling choices and the need for caution when comparing parameter estimates across waveform families, especially when interpreting claims of eccentricity or inferring astrophysical formation scenarios. The development of faithful eccentric and precessing waveform models that capture the full inspiral-merger-ringdown will be essential for robust inferences in future analyses, particularly to help distinguish potential degeneracies between precession and eccentricity.

While previous analyses have provided important constraints on the eccentricity of GW200105 and other NSBH mergers, they have been fundamentally limited by the capabilities of existing waveform models. For example, \textsc{TaylorF2Ecc} includes post-Newtonian (PN) eccentricity effects but is restricted to the inspiral regime and assumes aligned spins, omitting both higher-order PN corrections and precession~\citep{taylorf2ecc}. \textsc{PyEFPE}, a more recent model, incorporates both spin-induced orbital precession and eccentricity but remains limited to the inspiral phase and does not include merger or ringdown contributions \citep{pyEFPE}. In contrast, \textsc{IMRPhenomXP} and \textsc{IMRPhenomXPHM} model the full inspiral-merger-ringdown (IMR) evolution and include precession (with the latter adding HOMs), but both are restricted to quasicircular orbits and therefore cannot capture eccentricity \citep{IMRPhenomXP_1,IMRPhenomXP_2}. The absence of waveform models that fully capture both eccentricity and the complete IMR evolution—including precession and HOMs represent a major limitation of previous studies. The recently introduced \textsc{IMRPhenomTEHM} model attempts to bridge this gap by incorporating eccentric PN dynamics, HOM, and an accurate quasicircular limit, but it remains limited to aligned spins and is still undergoing validation \citep{IMRPhenomTEHM}. Notably,~\textsc{IMRPhenomTEHM} is comparable to \seob for low masses, mass ratios, and low eccentricity. However, exhibits relatively large mismatches when compared with \seob in certain regions of parameter space (larger eccentric values, HOM, large mass ratios) \citep{IMRPhenomTEHM}, reflecting the fact that, as a newly developed model, further work is required to improve its faithfulness across the full range of astrophysical scenarios.

\section{Result Verification}
We performed various checks to verify the eccentricity we measured in GW200105. First, we visually checked for the absence of glitches or non-stationary Gaussian noise in the data.  Such artifacts could produce spurious evidence for eccentricity if present, but we found no such anomalies.

Next, we verified the integrity of waveform generation at higher eccentricities, confirming that the waveforms were free of boundary effects and numerical irregularities. No abnormalities were observed. We also ensured that the waveforms were smoothly tapered to avoid sudden onsets that might mimic eccentric features. Additionally, all likelihood calculations were initialized 1 Hz above the lower frequency cutoff to guard against pseudo-eccentricity effects due to rigid waveform start times.

Previous data quality studies were performed by~\citet{2025arXiv250315393M}, who filtered their analysis at higher frequencies and found no impact on the measured eccentricity or any signs of late-time noise contamination. In this work, we extend the investigation to lower frequencies to examine whether low-frequency noise could artificially influence the inferred eccentricity. We performed tests by varying the starting frequency (18, 19, 20, and 21 Hz) while fixing the reference frequency at 20 Hz. As expected, lower starting frequencies yielded slightly higher measured eccentricities, consistent with the inclusion of additional early inspiral signal where eccentricity has a stronger imprint. No anomalous trends or artifacts were observed across these runs, and we found no evidence that low-frequency noise contributed spurious support for eccentricity.

To assess how much residual eccentricity remains in the less sensitive high-frequency band, we repeated the analysis while computing the likelihood above 30 Hz, and setting the reference and the lower frequency cutoff to 20 Hz. This effectively removes some of the eccentricity information in the early parts of the inspiral, where eccentricity leaves its strongest imprint. The inferred eccentricity remained relatively high at $0.124^{+0.027}_{-0.066}$, which could potentially be due to some underlying noise artifact. To validate this result, we performed a targeted injection study in which we created a fiducial signal with the maximum-likelihood parameters from the posteriors of the 30 Hz run. We inferred the source properties of this simulated signal with the same frequency settings as the initial test and recovered an eccentricity of $0.122^{+0.025}_{-0.026}$ consistent with the actual data. This suggests that such measurements are plausible and expected without invoking non-Gaussian noise.

One notable feature in the posterior distribution for GW200105 is the presence of two distinct islands in the eccentricity and chirp mass (or mass ratio) posteriors. This island was also observed by~\citet{2025arXiv250315393M}. To investigate this feature, we varied the low-frequency cutoff in the likelihood calculation. As the cutoff was increased from 20 Hz to 30 Hz, the islands in the posterior distribution merged into one final island. To further assess this effect, we selected a high-likelihood point from the 30 Hz posterior, which initially resided in the valley of the 20 Hz posterior distribution. We recalculated the likelihood at this point and observed an increase as the frequency cutoff was raised. This indicates that the eccentric waveform model effectively rules out posterior samples in the valley, and the appearance of the islands is not due to waveform generation errors, sampling artifacts, or other numerical issues. Furthermore, we also tested different samplers, including \textsc{Multinest}~\cite{2009MNRAS.398.1601F}, and the islands were also present in this run. Overall, we find that the islands are physical and not an effect of sampling or waveform errors. These islands may be due to the degeneracies between the chirp mass and eccentricities we discuss in Sec.~\ref{sec:challanges}. Overall, it is still unclear what is causing these two distinct islands.

The accuracy of \teob and \seob models are not well validated at larger mass ratios, which motivated our original mass ratio prior bound range of  $q \ \epsilon \ [1,6]$. To assess the impact of this choice, we performed a follow-up analysis with an expanded prior  $q \ \epsilon \ [1,15]$. Overall, we found that our initial measurements were consistent with the new run and our mass ratio agrees with what was measured in previous analysis \cite{planas2025eccentricinspiralmergerringdownanalysisneutron}.

To further test the preference for eccentricity, we calculated the Bayes factor using the Savage-Dickey density ratio (SDDR), which is useful for nested models~\citep{2021arXiv210513270T}. SDDR is given as:
\begin{equation}   
    \text{SSDR} = \frac{p(d \mid h_1)}{p(d \mid h_0)}  = \frac{p(e \mid d, h_1)}{p(e=0 \mid h_1)}\,,
\end{equation}
\noindent
where, $p(d \mid h_0)$ is the marginal likelihood evaluated for a quasi-circular model $h_0$ and $p(d \mid h_1)$ for the eccentric model $h_1$ model respectively. This ratio provides a direct measure of the Bayes Factor in favor of the eccentricity hypothesis, with values greater than 1 indicating preference for a non-zero eccentricity model.

We computed the Bayes factor in favor of the eccentric model over the non-eccentric model by evaluating the ratio of the posterior to the prior density evaluated at $e=0$. This yielded a Bayes factor of approximately \bfseob for GW200105 using \seob and \bfteob using \teob. For the HOM, we obtain the Bayes factor to be \bfseobHM and \bfteobHM for \seob and \teob, respectively, for a bound ranging from 0 to 0.2 on eccentricity. This indicates roughly $6-7$ times more preference for the eccentric model over the non-eccentric one, assuming a uniform prior. We also calculate the Bayes factor for a uniform log prior for a dominant mode \teob model, and we find the value to be \bflogteob in favor of the eccentric model with a bound ranging from $10^{-4}$ to 0.2. The Bayes factors calculated from SDDR are consistent with the evidence reported in \textsc{Dynesty}. While there is still slight support from the uniform log prior run, the evidence is not robust enough to conclusively identify GW200105 as an eccentric binary.

\begin{figure*}[ht!]
    \centering
    \includegraphics[width=0.47\textwidth]{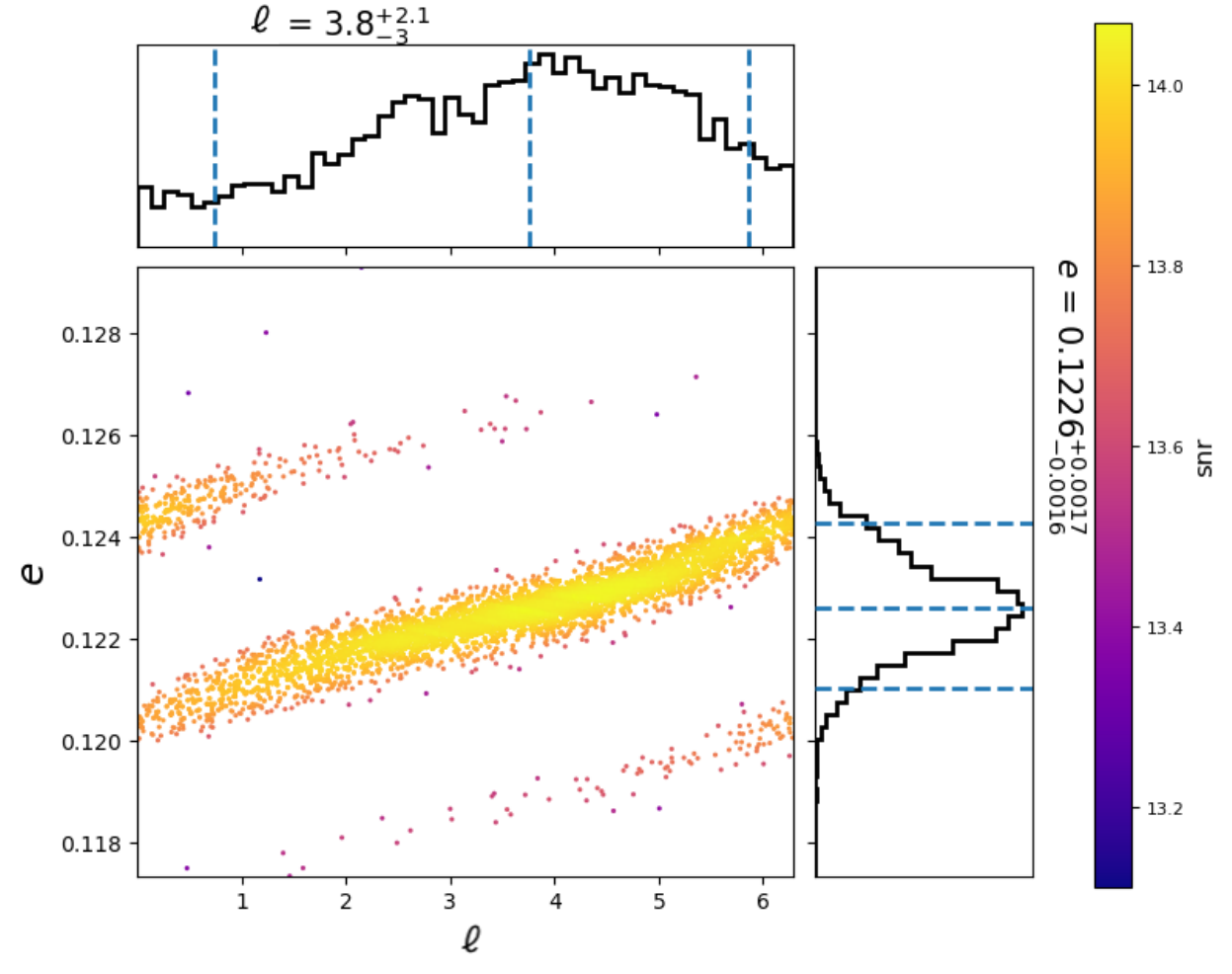} 
    \includegraphics[width=0.47\textwidth]{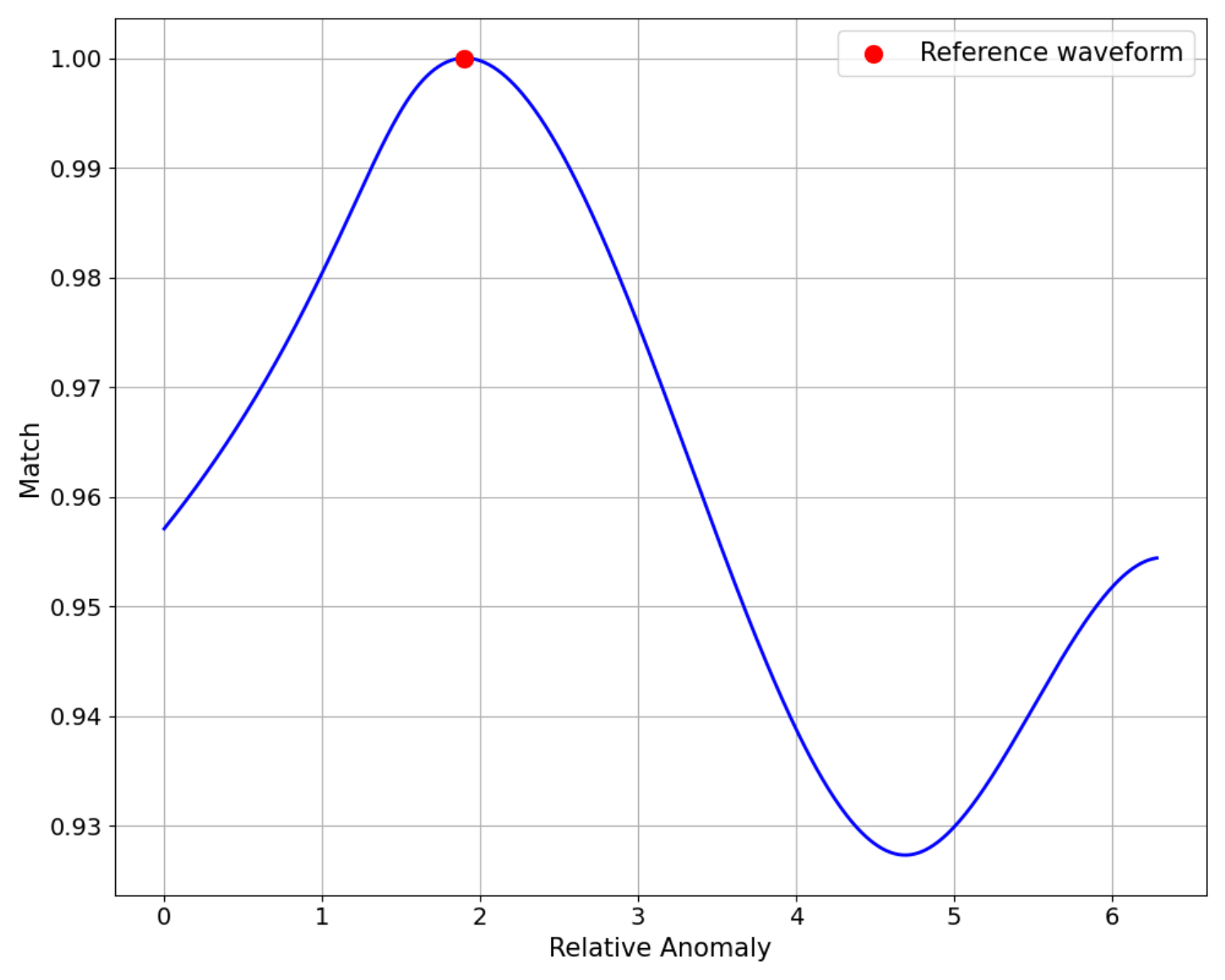} \vspace{-0.6mm}
    \includegraphics[width=0.47\textwidth]{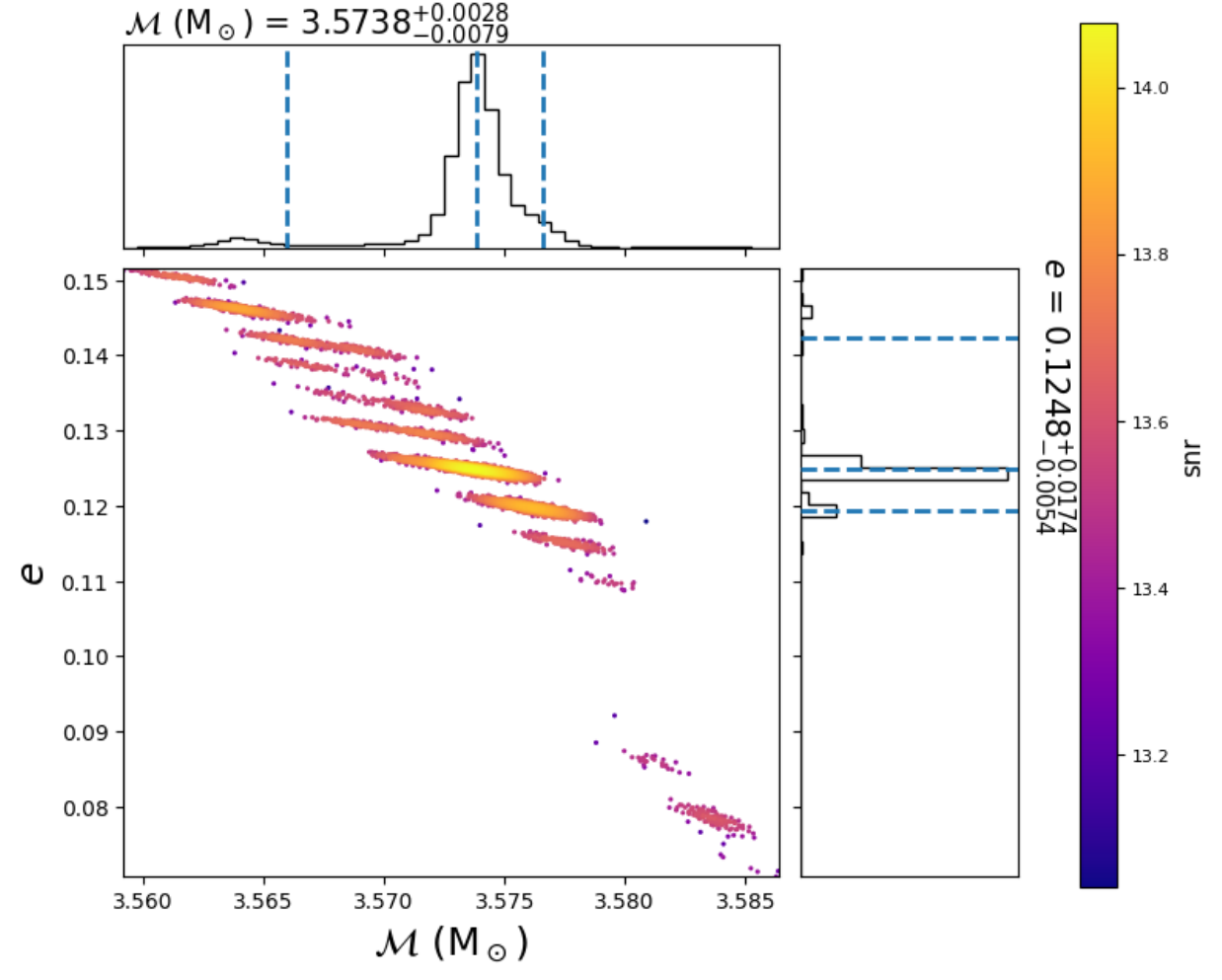}
    \includegraphics[width=0.47\textwidth]{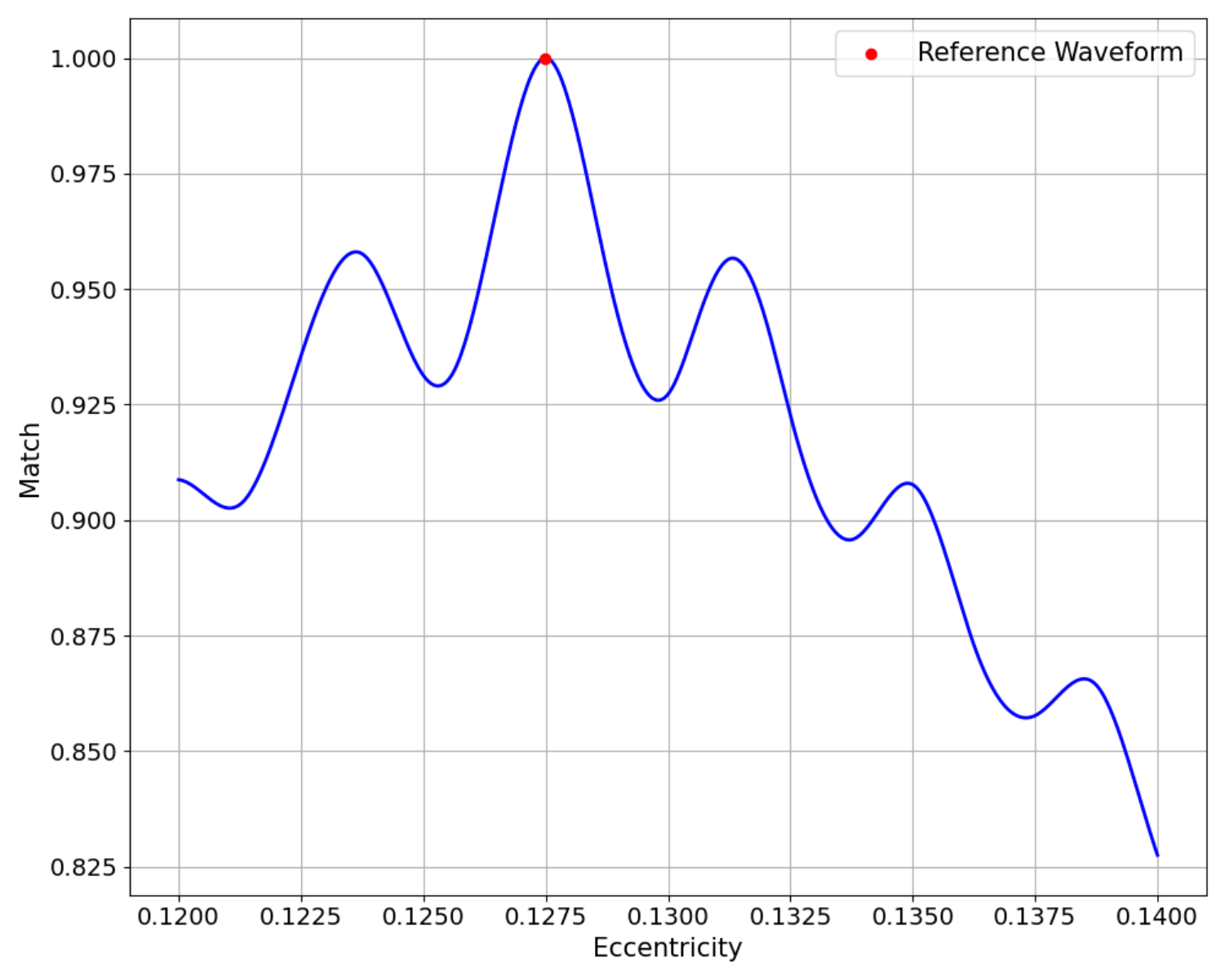} \vspace{-0.6mm}
    \caption{Marginalized posterior distributions and matches for two configurations: one with only eccentricity and anomaly varied (top), and the second with chirp mass and eccentricity varied (bottom). The model used for the plot is the dominant (2,2) mode \seob. The same test was performed for \teob and the same distribution is present in both waveforms. Multiple distinct modes are visible in parameters such as eccentricity, anomaly, and masses, illustrating the complex structure of the likelihood surface. The left plots show the inferred posteriors from parameter estimation, while the right plots show waveform matches for a fixed reference waveform depicted with the red point. The multi-modal structure in parameter space reflects underlying features of the waveform evolution.}
    \label{fig:modes}
\end{figure*}

\subsection{Parameter Estimation Challenges} \label{sec:challanges}

We encountered several challenges when using the \seob and \teob waveform models, most notably the presence of a complex multi-modal structure in the likelihood. These arise, for example, when varying the relativistic anomaly in \seob or the true anomaly in \teob, and are also evident along mass parameters. Crucially, these features are not artifacts of undersampling but rather originate from the waveform models themselves. This can be demonstrated by computing the match between a fixed reference waveform and a set of waveforms with identical parameters except for eccentricity. The resulting match curve shows clear repeated peaks corresponding to the distinct modes observed in Fig.~\ref{fig:modes}. This strongly suggests that the features are imprinted directly in the waveform as a function of eccentricity or anomalies. The physical origin of these structures remains unclear and presents an interesting direction for future study. These complexities also complicate the inference process, as illustrated in Fig.~\ref{fig:modes}, where even with most parameters fixed, the posterior remains multi-modal and will be further complicated when the full parameter space is explored.

\section{Conclusion and Discussion}

In this work, we conducted the first comprehensive analysis of eccentricity in seven confidently detected low-mass gravitational-wave events reported by the LIGO–Virgo–KAGRA (LVK) collaboration to date using a newly developed waveform model \seob. To assess the impact of waveform systematics, we cross-validate our results using the \teob model. 

Our findings reinforce the expectation that most low-mass binaries observed to date are consistent with circular inspirals (see Table~\ref{tab:results}). We also placed the first upper bound on the orbital eccentricity, ranging from $e<0.063$ to $e<0.075$ for GW190814, and found no support for eccentricity in this event. Although the inclusion of HOMs slightly reduces the eccentricity estimates in GW190814, the results remain broadly consistent across waveform models. In contrast, GW200105 shows moderate evidence for non-zero eccentricity, with constraints of $e<0.14$ at a 90\% confidence interval when eccentricity is sampled uniformly between 0 and 0.2. This result persists across both waveform models and inference settings and remains consistent under a range of validation tests, including variations in the frequency cutoff, sampling diagnostics, and waveform tapering procedures. However, we find that the eccentricity constraint is sensitive to the choice of prior, as also noted in Refs.~\cite{2025arXiv250315393M,planas2025eccentricinspiralmergerringdownanalysisneutron}. When repeating the analysis using a prior uniform in $\log_{10}(e)$ with a lower bound of $10^{-4}$, we found that the support for moderate eccentricity is significantly reduced to twice in favor of eccentricity. This demonstrates that prior assumptions can significantly influence the inferred eccentricity distribution, particularly when the likelihood surface is broad or multi-modal.

We ensure that the observed eccentricity in GW200105 is not an artifact of detector noise, waveform systematics, or prior assumptions by performing several verification tests. First, we visually inspected the data for glitches or signs of non-stationary Gaussian noise and found no anomalies that could mimic eccentricity. We then verified that our eccentric waveform models produced stable and well-behaved signals at higher eccentricities, with no signs of tapering issues or boundary artifacts. To test the robustness of the inference, we varied the low-frequency cutoff from 18 to 21 Hz while keeping the reference frequency fixed. The recovered eccentricity remained non-zero across these configurations and increased slightly at lower cutoffs, consistent with improved sensitivity to early inspiral features, and no lower noise effects were present. Finally, we tested the loss in eccentricity information by inferring the eccentricity using 30 Hz as the starting frequency of our likelihood calculation and a reference lower frequency of 20 Hz for waveform generation. The eccentricity remained relatively high, and we verified this result by generating an eccentric injection at the maximum-likelihood parameters from the original 30 Hz test. Our injection recovery was consistent with the eccentricity measured in the test. These cross-checks affirm that the measured eccentricity in GW200105 is not a numerical or modeling artifact, but a robust feature supported by multiple lines of evidence.

Despite these findings, several factors highlight the need for continued investigation into the nature of eccentricity in compact binary mergers. One concern is the potential degeneracy between eccentricity and spin-induced precession. Although this effect is less significant for low-mass systems like GW200105~\citep{nsbh_ligo,2025arXiv250315393M}, it demands further study with improved waveform models and high signal-to-noise ratio events. Additionally, GW200105 was observed by only one detector, which limits our ability to fully rule out noise-induced features that may mimic eccentricity. Nevertheless, our injection studies performed with a starting frequency of 30 Hz recover eccentricities consistent with the injected values, making it unlikely that noise alone accounts for the observed signal. Finally, we also identified a secondary mode or island in the posterior distribution which involves a trade-off between the chirp mass and eccentricity, consistent with the findings in \citet{2025arXiv250315393M}. A bimodality near $e \sim 0.12$ in the main peak, previously reported by~\citet{planas2025eccentricinspiralmergerringdownanalysisneutron}, is also recovered in our analysis. Interestingly, while~\citet{2025arXiv250315393M} and~\citet{planas2025eccentricinspiralmergerringdownanalysisneutron} report slight disagreements between their two runs regarding the number and locations of posterior peaks, our analysis recovers all of these features with both waveform models. This suggests that, when considered together, the two prior studies are mutually consistent and that our results are compatible with both. The presence of these degeneracies underscores the complexity of parameter estimation in such systems, particularly when multiple physical effects persist.

The possibility of non-zero eccentricity in GW200105 carries significant astrophysical implications. BBH mergers are often linked to dynamical formation in dense environments such as globular clusters~\citep{Wen:2002km,romeroshaw2025gw200208222617eccentricblackholebinary} and AGN disks~\citep{Romero-Shaw:2022xko}, where repeated interactions can induce orbital eccentricity. In contrast, NSBH systems like GW200105 have traditionally been associated with isolated binary evolution. The evidence for eccentricity in GW200105 challenges this assumption and points to a dynamical origin for at least some NSBH mergers. However, the specific dynamical channels responsible may differ from those relevant to BBHs. The tidal susceptibility of neutron stars and their low masses make them less likely to survive repeated interactions in dense stellar environments~\citep{dhurkunde_nitz_2025}. Instead, hierarchical triple systems or interactions in nuclear star clusters offer more plausible formation scenarios for eccentric NSBH binaries~\citep{Mandel:2021smh}. Although globular clusters are prominent sites of BBH mergers, their ability to produce and retain NSBH systems remains uncertain, introducing potential tension with that channel for such events. 

As the number of NSBH detections increases, so does the potential to verify formation channels that produce eccentric NSBH populations. The fourth observing run of the LVK collaboration has so far yielded approximately 300 low-latency alerts, including several NSBH candidates. With continued observations, these formation channels may soon be confirmed~\citep{Akyuz_2025ype,gracdb_dcc}.  As the current detector network grows in sensitivity and detection rate, next-generation observatories such as Cosmic Explorer~\citep{2019BAAS...51g..35R,2023arXiv230613745E} and the Einstein Telescope~\citep{2010CQGra..27a5003H,2010CQGra..27s4002P} are expected to observe thousands of mergers annually. This rapid growth will place increased demands on parameter estimation pipelines, particularly when using waveform models that include features like eccentricity, spin precession, and higher-order modes. These models are computationally expensive and require dense sampling of multi-modal parameter spaces in parameters such as mass, eccentricity, and anomaly.

To make full use of the data provided by future detectors, more efficient inference techniques and waveform models are essential. In our analyses, the dominant computational cost arose from waveform generation. Reducing the waveform generation time is therefore a key priority. ROMs offer a promising route to accelerate models such as \seob and \teob, making them a critical focus for future development. 

Another key strategy for computational efficiency involves marginalization schemes that reduce sampling complexity by integrating over nuisance parameters. For instance, marginalizing over phase and anomaly could help mitigate multi-modality in the likelihood. Inference runs that included higher-order modes were approximately three times slower than those restricted to the dominant mode, primarily due to the inability to marginalize over the coalescence phase compared to our (2,2) mode runs. Future developments in marginalization methods, such as the phase marginalization for higher order harmonics demonstrated in \citet{importance_sampling}, could improve performance, especially for low-mass systems, where these modes are more influential. 

In this work, we have already experimented with marginalization over the total mass to avoid sampling over the modes and help reduce the number of waveform generations. The model we implemented takes an initial waveform and stretches or squeezes it to achieve the desired final total mass. This can help avoid having to generate multiple waveforms for a set of given parameters. We have not tested the model for the longest duration signals, but it may be beneficial towards BNS systems where waveform generation takes $\ge 10$ seconds. In practice, we found the method to be inefficient for GW200105 and chose not to apply it in our final analyses.

Understanding the degeneracy structure between anomaly, eccentricity, and mass could also lead to reparameterization using a pseudo-anomaly or effective eccentricity. Such a transformation may help smooth over the discrete modes in the likelihood, effectively turning the distribution Gaussian and improving sampling efficiency. If the structured bands observed in waveform matches reflect unresolved modes, then designing samplers that can efficiently resolve these narrow likelihood features will be crucial. These advances will be necessary to enable fast and robust eccentricity measurements in the high-throughput era of the Cosmic Explorer and the Einstein Telescope, and the anticipation of the O4 data release.

The necessary information to reproduce these results and the posterior files can be found on GitHub \citep{ecc_pe_repo}

\begin{acknowledgments} 

KK, KS, and AHN acknowledge support from the NSF grant PHY-2309240. The authors acknowledge the support from Syracuse University for providing computational resources through the OrangeGrid High Throughput Computing (HTC) cluster supported by the NSF award ACI-1341006. The authors express their sincere gratitude to Rossella Gamba for her invaluable assistance and for providing access to \textsc{TEOBresumS}, which greatly facilitated this work. KK also extends thanks to Yifan Wang for developing the plugin used to generate the gravitational waveforms in \textsc{PyCBC} and for the many helpful discussions, comments, and suggestions that contributed to this project. Finally, we are deeply grateful to Aldo Gamboa and Alessandra Buonanno for permitting the use of the \seob eccentric waveform model, which was integral to our analysis, and for the feedback received on the paper. Additionally, KK would like to thank Aldo Gamboa for all the discussions related to the modes discovered in the eccentric waveforms. KK would like to thank Gonzalo Morras for allowing us to use their posteriors in our plots, and Maria de Lluc Planas et al. for making their posteriors publicly available. The authors thank Jenny Vo and Paw Eh Blut Say, high school students who initially performed a preliminary analysis to investigate the eccentricity in GW190814 through the Syracuse University Research in Physics (SURPh) program. Their early contributions helped lay the groundwork for what evolved into a broader investigation. 
 
This research has used data or software obtained from the Gravitational Wave Open Science Center \cite{GWOSC1,GWOSC2,gwosclink}, a service of the LIGO Scientific Collaboration, the Virgo Collaboration, and KAGRA. This material is based upon work supported by NSF's LIGO Laboratory which is a major facility fully funded by the National Science Foundation, as well as the Science and Technology Facilities Council (STFC) of the United Kingdom, the Max-Planck-Society (MPS), and the State of Niedersachsen/Germany for support of the construction of Advanced LIGO and construction and operation of the GEO600 detector. Additional support for Advanced LIGO was provided by the Australian Research Council. Virgo is funded, through the European Gravitational Observatory (EGO), by the French Centre National de Recherche Scientifique (CNRS), the Italian Istituto Nazionale di Fisica Nucleare (INFN) and the Dutch Nikhef, with contributions by institutions from Belgium, Germany, Greece, Hungary, Ireland, Japan, Monaco, Poland, Portugal, Spain. KAGRA is supported by Ministry of Education, Culture, Sports, Science and Technology (MEXT), Japan Society for the Promotion of Science (JSPS) in Japan; National Research Foundation (NRF) and Ministry of Science and ICT (MSIT) in Korea; Academia Sinica (AS) and National Science and Technology Council (NSTC) in Taiwan.

\end{acknowledgments}

\bibliography{references}

\end{document}